\documentstyle[epsfig,twocolumn,pra,aps]{revtex}
\begin{document}
\draft

\wideabs{

\title{Asymmetric quantum channel for quantum teleportation}

\author{M. S. Kim$^1$ and Jinhyoung Lee$^{1,2}$}

\address{$^1$ School of Mathematics and Physics, The Queen's
  University of Belfast, \\ BT7 1NN, United Kingdom\\
$^2$ Department of Physics, Sogang University, CPO Box 1142, Seoul, Korea}

\date{\today} 

\maketitle

\begin{abstract}  
  We propose a realistic optimal strategy for continuous-variable teleportation
  in a realistic situation.   We show that any imperfect quantum operation
  can be understood by a
  combination of an {\it asymmetrically-decohered} quantum channel and
  perfect apparatuses for other operations.  For the
  asymmetrically-decohered quantum channel, we find some
  counter-intuitive results; teleportation does not
  necessarily get better as the channel is initially squeezed more.
  We show that decoherence-assisted measurement and transformation may
  enhance the fidelity for the asymmetrically mixed quantum channel.
\end{abstract}


}

\newpage 

Quantum teleportation is one of the important manifestations
of quantum mechanics.  In particular, quantum teleportation of
continuous variable states \cite{Vaidman94} has attracted a great deal
of attention because of a high detection efficiency, handy
manipulation of continuous variable states
\cite{Braunstein98,Ralph98,Braunstein98-2}, and possibility of
application to high-quality quantum communication.  
Two kinds of protocols have been suggested for
continuous variable teleportation; one utilizes the entanglement
between quadrature-phase variables \cite{Braunstein98} and the other
between the photon-number sum and the relative phase \cite{Milburn99}.
Both the protocols employ a squeezed two-mode vacuum for the quantum
channel.  In this paper, we report how to optimize the quantum
teleportation of quadrature-phase variables when the quantum channel
and experimental conditions are not perfect.

There are a few obstacles which make the teleportation of
quadrature-phase variables imperfect.  The perfect quantum
teleportation is possible only by a maximally-entangled quantum
channel, {\it i.e.}, by an infinitely squeezed state which is
unphysical as it incurs the infinite energy.  Moreover, when the
quantum channel is exposed to the real world, it is influenced by the
environment, which turns the {\it pure} squeezed state into a {\it
  mixture} and deteriorates the entanglement property.  To maximize
the channel entanglement, purification protocols for continuous
variable states have been suggested by Parker {\it et al.}
\cite{Parker00} for partially-entangled pure states and by Duan {\it
  et al.} \cite{Duan99-1} for mixed Gaussian states.  
However, the theoretical suggestions have not been
realized by experiment.  Further, there are other obstacles in
experiment such as imperfect detection efficiency at the sending
station and imperfect unitary transformation at the receiving station.
We show that the imperfect conditions may be absorbed into the
imperfect quantum channel while other apparatuses are treated perfect,
and find the optimization condition for the teleportation under a
given experimental condition.  We show that blindly maximizing the
initial entanglement of the quantum channel does not necessarily bring
about the best teleportation.

Two modes $a$ and $b$ of the squeezed vacuum are distributed,
respectively, to a sending and a receiving station. 
At the sending station, the original unknown
state is entangled with the field mode $a$ of the quantum channel by a
50/50 beam splitter.  Two conjugate quadrature variables are measured
respectively for the two output fields of the beam splitter using
homodyne detectors.  Upon receiving the measurement results through
the classical channel, the other mode $b$ of the
squeezed vacuum is displaced accordingly at the receiving station.
The quantum teleportation of quadrature-phase variables is well described
by a phase-space distribution, in particular, the Wigner function \cite{Wigner}
that is the Fourier transform of its characteristic function $C(\eta)\equiv
\mbox{Tr}[\rho\hat{D}(\eta)]$ for the state of the density operator $\rho$.
$\hat{D}(\eta)$ is the displacement operator.   

The quantum teleportation is completed
by a unitary displacement operation at the receiving station.  If a
field state of its Wigner function $W(\alpha)$ is displaced by
$\beta$, it is represented by the Wigner function $W(\alpha-\beta)$.
In the experiment, the displacement operation is performed using a
beam splitter of a high transmittance $T$ \cite{Ralph98,Furusawa98}.
To displace a field state of the Wigner function $W(\alpha)$, it is
injected into the beam splitter while a high intensity coherent state
of amplitude $\beta/\sqrt{1-T}$ is injected to the other input port.
The beam splitter operation results in the convolution between the
two input states \cite{Kim95}.  Remembering that the synthesizing coherent 
state is the displaced vacuum, we find the Wigner function 
$W_d(\gamma-\beta)$ of the output field by
\begin{equation}
\label{convolution-relation-2}
W_d(\gamma-\beta)={1 \over 1-T} \int d^2\alpha W(\alpha)W_{vac}
\left({\gamma -\beta-\sqrt{T}\alpha\over \sqrt{1-T}}\right)
\end{equation}
where $W_{vac}(\alpha)$ is the Wigner function for the vacuum.  We can
easily see that {\it displacing a field by a beam splitter of its
  transmittance $T$ is equivalent to unitarily-displacing the field
  after it is mixed with the vacuum at a beam splitter of the same
  $T$}.  Note that mixing a field with the vacuum at a beam splitter
results in the same dynamics of the field influenced by the vacuum
environment.  Kim and Imoto \cite{Kim95} found that assuming the coupling 
of the system with the environment $\kappa$ 
and the exposure time to the environment $\tau$, the normalized
interaction time $R\equiv 1-\exp(-\kappa\tau)$ is the same as $1-T$.

The inefficient detection at the sending station is another factor
which degrades the teleportation \cite{Braunstein98,Ralph99}.  When
the two photomultipliers of a homodyne detector have the same
efficiency $\eta$, the imperfect homodyne detector is described by a
perfect homodyne detector with a beam splitter in front
\cite{Leonhardt94}.  A field passes through the beam splitter of the
transmittance $\eta$ and it is mixed with the vacuum which has been
injected into the other input port.  The inefficiency of the detection
can also be passed to the quantum channel.  We will discuss later that
inefficiency at the sending station gives an effect not only to the
quantum channel but also to the original unknown field to teleport.

We have seen that imperfect operations at the receiving and sending
stations can be understood as a combination of the perfect operations
with an imperfect mixed quantum channel.  
Imperfection at the displacement operation is absorbed by the field
mode to the receiving station.  The field mode to the sending station
can absorb inefficiency in the homodyne detection.  These
considerations lead the quantum channel mixed {\it asymmetrically} due to a
different condition for each mode of the quantum channel.  
The study of the teleportation using the {\it asymmetrically-decohered} quantum channel 
 is important not only because it
can explain the experimental situation but also because it gives novel
features and deeper understanding of the nature of entanglement for 
the continuous variables.  If quantum
teleportation is used for a quantum communication, it is more likely
that the two modes of the quantum channel will undergo different
environmental conditions.  To the best of our knowledge, the impact of
the asymmetric channel on the quantum teleportation has not yet been
seriously explored.

The quantum channel, which is initially in the two-mode squeezed
vacuum of squeezing factor $s$, is influenced by the thermal
environments.  Assuming that two thermal modes are 
independently coupled with the quantum channel, the dynamics of the 
quantum channel is described by a Fokker-Planck equation in the interaction
picture \cite{Jeong00}.  Solving the equation, the time-dependent Wigner
function is obtained as
\begin{eqnarray}
\label{quantum-channel}
W_{ab}(\alpha_a,\alpha_b)={\cal N}\exp\bigg\{&-&{2 \over m_a m_b -c_a c_b}
\big[m_a|\alpha_a|^2+m_b|\alpha_b|^2 \nonumber\\
&+&\sqrt{c_a c_b}(\alpha_a\alpha_b+\alpha_a^*\alpha_b^*)\big]\bigg\},
\end{eqnarray}
where $m_i=R_i(1+2\bar n_i)+T_i\cosh 2s$ and $c_i=R_i\sinh 2s$,
($i=a,b$); $\bar n_i$ is the average thermal photon number of the
environment for the channel mode $i=a,b$.  The normalized interaction
time $R_i(\equiv 1-T_i)$ is zero when the quantum channel is not
subject to the environment and grows to unity when the channel
completely assimilates the environment.  

It has been shown that a two-mode Gaussian state is separable when a
semi-positive well-defined $P$ function can be assigned to it after
some local operations \cite{Duan99,Lee00}.  The two-mode squeezed state
subject to the thermal environment is separable when
\begin{equation}
\label{separation-condition}
(m_a-1)(m_b-1)\geq c_a c_b.
\end{equation}
As a special case, if the channel mode $b$ is influenced by the
vacuum, {\it i.e.}, $\bar{n}_b=0$, the channel becomes separable when
\begin{equation}
\label{separation-conditon-2}
R_a\geq {1 \over 1 + \bar n_a}.
\end{equation}
For this case, the separation
condition depends only on the average thermal photons influencing the
channel mode $a$, even when the channel is minimally squeezed at the
initial instance.

By the ideal teleportation, the original state is recovered at the 
receiving station.  However, when the channel is not maximally entangled
the teleportation is not ideal.  The fidelity 
${\cal F}$ is defined as ${\cal F} = \pi\int
d^2\alpha W_o(\alpha)W_r(\alpha)$, where $W_o(\alpha)$ and 
$W_r(\alpha)$ are the Wigner functions, respectively, for the original and
teleported states, to show how close the teleported state is to the original state
\cite{Lee00}.  One of the important assumptions in teleportation is that
the original state is unknown so that the teleportation protocol has to be 
selected to optimize the average fidelity over all the possible original states.
However, the average of the fidelity defined above is zero for continuous variables
because of the scope of the possible original states \cite{Braunstein-jmo}.  
We thus take a subset composed of all coherent states and find the strategy
to optimize the teleportation. 
A coherent state is one of a few manifestly quantum and extremely useful states 
generated in laboratories.  Because the coherent states are nonorthogonal it is impossible
to discern them with certainty.  Any state can be written as a weighted sum of 
coherent states.

Let us consider the teleportation using the quantum channel
(\ref{quantum-channel}).  Before any action at the receiving and
sending stations, the total state is a product of the original unknown
state and the quantum channel.  
Setting homodyne detectors at the two output ports of the beam
splitter, quadrature-phase variables $p_1$ and $q_2$ are measured at
the respective output ports.  
Upon receiving each pair of measurement results
$g\equiv \sqrt{2}(q_2-i p_1)$, the quadrature
variables of the channel field $b$ is displaced by $g'(g)$ accordingly.
For a coherent original state $|\alpha\rangle$, we can calculate the 
measurement-conditioned Wigner function for the teleported state.
Using the definition of the fidelity given above, ${\cal F}(\alpha, g, g')$ is
calculated.  The average fidelity
for the total set of coherent states is found:
\begin{equation}
\label{fidelity}
\bar{\cal F}={\int d^2 \alpha \int d^2 g {\cal F}(\alpha, g, g') \over \int d^2\alpha}
= {\cal F}_o{\int d^2 g \exp[-{\cal E}|g-g'|^2] \over \int d^2\alpha}
\end{equation}
where ${\cal E}$ is a channel-dependent factor and
\begin{equation}
\label{fidelity-o}
{\cal F}_o ={1 \over 1 +{1 \over 2}(m_a+m_b)-\sqrt{c_a c_b}}.
\end{equation}
Eq.(\ref{fidelity}) readily shows that only when the displacement $g'$ is the same as
$g$, the average fidelity may have a finite value and $\bar {\cal F}={\cal F}_o$ is the optimum
average fidelity in this case. Braunstein and Kimble
found that the teleportation is optimized when $g=g'$ for the pure quantum channel 
and the fidelity ${\cal F}_o$ is greater than 1/2 when the pure channel is entangled \cite{Braunstein98}. 
Using Eqs.~(\ref{separation-condition}) and (\ref{fidelity-o}), we find that 
for the symmetrically decohered {\it mixed channel}, {\it i.e.}, $m_a=m_b$, 
the fidelity is greater than
1/2 as far as the channel is entangled.  However, when the channel is asymmetrically
decohered, the value 1/2 is not necessarily the critical fidelity for the standard teleportation
scheme described above.   Another important fact is found that
{\it the teleportation does not necessarily get
  better as the quantum channel is initially squeezed more}.  
To illustrate more clearly, assume $\bar n_b=0$.  For $R_a=0$ and $R_b=1$,
the quantum channel is inseparable as shown in 
Eq.(\ref{separation-conditon-2}) but Eq.(\ref{fidelity})
gives the fidelity ${\cal F}_o=1/(2+\sinh^2 s)$, which means that the more initially squeezed
the quantum channel, the smaller the fidelity is. 
 
In Fig.~1, the fidelity ${\cal F}_o$ is
plotted against initial squeezing $s$ for the asymmetric quantum
channel where the channel mode $b$ is influenced by the vacuum
environment for the normalized interaction time $R_b=0.01$ and 0.05
while the channel mode $a$ is not influenced by an environment.  The
teleportation via the quantum channel, which has been decohered by the
vacuum with interaction time $R_b=0.01$, in fact, corresponds to the
teleportation with the {\it pure} squeezed quantum channel and
imperfect displacement using the beam splitter of transmittance
$T=99\%$.  It is clearly seen that even for the quantum channel of
seemingly-negligible asymmetry, if the channel is initially squeezed
more than a certain degree, the teleportation becomes worse.  
By the first-derivative of ${\cal F}_o$ with regard to $s$, we find that the
teleportation is optimized when the squeezing is
$\mbox{e}^{-2s}=|t_a-t_b|/ (t_a + t_b)$
for a fixed channel condition $t_a$ and $t_b$, where $t_i=\sqrt{T_i}$ ($i=a,b$).  Note that this result
does not depend on the temperature of the environments.  

For a mixed quantum channel, is the standard teleportation of the orthogonal
measurement and the unitary transformation the best strategy?  The straightforward answer is
beyond the scope of this paper but the following argument gives some hint.  Let us consider the
asymmetrically mixed channel by taking $\bar n_b=0$ and $\bar n_a\neq 0$.  For given initial squeezing
and $T_a$, the fidelity is maximized to ${\cal F}_o=1/[1+(1+\bar n_a)(1-T_a)]$ only when 
$T_b=(\sinh2s/2\sinh^2 s)^2 T_a$.  We find that {\it the critical fidelity 1/2 is recovered} to coincide with
the separation condition under this condition. This shows that some further decoherence may enhance 
the fidelity \cite{Ho}.  The transformation accompanied by decoherence is no longer unitary which implies
that a general transformation may optimize the fidelity.  
With the same argument, we find that a general measurement
may optimize the fidelity for asymmetrically mixed quantum channel.

Why does it happen?  One may {\it intuitively} expect that the more the
quantum channel is squeezed, the better teleportation is.  It is true
when the channel remains {\it pure} and not influenced by an
environment.  However, when the channel is {\it asymmetrically mixed}, the conjecture may be wrong.
There are many parameters which influences the teleportation.  To make
the analysis simple without losing the interesting features, we assume
for the rest of the paper that the channel is exposed to
low-temperature environments only for short periods of time, {\it
  i.e.}, $R_i \ll T_i /\bar{n}_i$.  The separation condition
(\ref{separation-condition}) shows that {\it the quantum
  channel remains entangled longer as it is initially squeezed more}.
However, as we have seen in Fig.~1, the fidelity of teleportation can
be worse with increasing the initial squeezing.  For the short interaction time with
low-temperature environments, we write the
Wigner function (\ref{quantum-channel}) to highlight the EPR
correlation \cite{EPR35} of the quantum channel as follows
\begin{eqnarray}
\label{quantum-channel-EPR}
W_{ab}(\alpha_a,\alpha_b)&\approx & {\cal N}\exp\Big({-2 \over m_a m_b
  - c_a c_b}\Big[ 
\mbox{e}^{2s}\{(t_b q_a +t_a q_b)^2 + t_b p_a
-t_a p_b)^2 \} 
\nonumber \\
&& +\mbox{e}^{-2s}\{(t_b q_a - t_a q_b)^2 + (t_b p_a
+t_a p_b)^2 \}\Big]\Big). 
\end{eqnarray}
As the initial squeezing
$s\rightarrow\infty$, the Wigner function $W_{ab}\rightarrow
C~\delta(t_b q_a + t_a q_b)\delta(t_b p_a - t_a p_b)$.
It is clear that the asymmetric channel has the EPR correlation
between the scaled quadrature variables; positions $q_a$ and
$q_b'=(t_a/t_b) q_b$ and momenta $p_a$ and
$p_b'=(t_a/t_b) p_b$.  This is somewhat similar to the relation
between an original phase space and amplified or dissipated phase
space depending on the scale factor $t_a/t_b$.  When the channel is
strongly squeezed the channel entanglement teleports the original
photon to the scaled space, which brings about large noise in the
teleported state.  For a large squeezing, even though the quantum
channel is strongly entangled, the strong entanglement results in
inefficient teleportation because the entanglement is between the
scaled spaces.

The fact that the asymmetrically mixed quantum channel shows quantum
correlation in differently-scaled spaces is not inherent in the two-dimensional qubit 
system where maximally correlated observables remain unchanged under the 
asymmetric decoherence.  Consider a bipartite system in a spin-singlet state which is asymmetrically
decohered in the phase-insensitive and isotropic environment as for the continuous variable system
we discuss in this paper.
Its correlation function for the spin variables along the directions {\bf a}
and {\bf b} of two subsystems is given by $\langle \hat{U}({\bf a})\hat{\sigma}_z
\hat{U}^\dag({\bf a})\otimes\hat{U}({\bf b})\hat{\sigma}_z\hat{U}^\dag({\bf b})\rangle
\propto -\cos\theta_{ab}$, where $\hat{U}$ is a unitary operator and $\theta_{ab}$  the
relative angle of the two vectors {\bf a} and {\bf b}.  We clearly see that the maximal
correlation between anti-parallel is preserved even after the asymmetric decoherence.   

An imperfect detection efficiency at the receiving station can also be
analyzed as a combination of perfect detection with a beam splitter in
front.  At the beam splitter, not only the channel state but also the
unknown original field are mixed with the vacuum.  This is why the average 
fidelity is maximized to ${\cal F}= 1/[\mbox{e}^{-2s}+1/T_a]$ for the
displacement factor $g'=g/t_a$ assuming the channel mode $b$ is not subject
to the environment.  
This shows that the teleportation is more
efficient when the channel is initially squeezed more, which is in
agreement with the earlier result \cite{Braunstein98}.

We have shown that the important experimental errors can be absorbed
by an imperfect mixed quantum channel while the experimental
operations are assumed to be perfect.  Because the experimental error
does not occur symmetrically between the receiving and sending
stations, it is important to study the influence of the
asymmetrically-decohered quantum channel on the teleportation.  
For the asymmetrically-decohered quantum channel we found
that the strong initial squeezing does not always optimize the
teleportation because the asymmetric quantum channel has the EPR
correlation between the differently-scaled phase spaces.  We found that a  
measurement and transformation accompanied by decoherence
optimizes the teleportation.

\acknowledgements

This work
was supported in part by the Brain Korea project (D-0055) of the
Korean Ministry of Education.

\begin{figure}
  \begin{center}
    \includegraphics[width=0.53\textwidth]{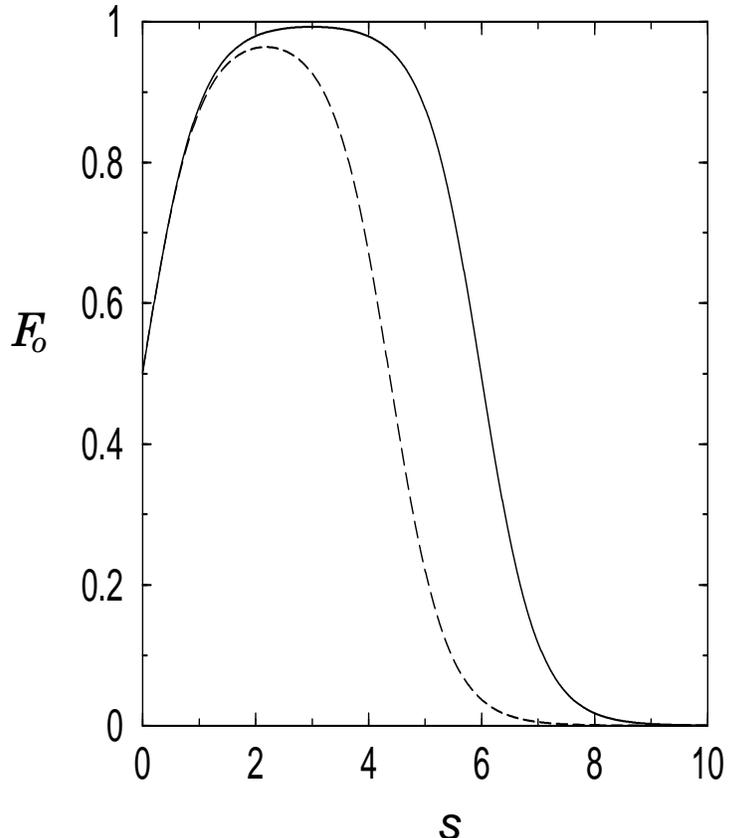}
    \caption{The fidelity ${\cal F}$
      is plotted against the degree of initial squeezing $s$.  The
      channel mode $a$ is not subject to the environment, {\it i.e.},
      $R_a=0$ and the channel mode $b$ is subjected to the vacuum
      environment with the normalized interaction time $R_b=0.01$
      (solid line) and 0.05 (dotted line).  The channel is thus
      inseparable.  The decohered quantum channel with $R_b=0.01$,
      corresponds to the {\it pure} channel for the imperfect
      displacement with the transmittance $T=99\%$.  }
    \label{fig:configuration}
  \end{center}
\end{figure}

\end{document}